\documentclass[jorunal]{IEEEtran}

\usepackage{amsmath}
\usepackage{amssymb}
\usepackage{amsthm}
\usepackage{xspace}

\usepackage[normalem]{ulem}	%for strikeout only
\usepackage{graphicx} % for windows
\usepackage{rotating} %for the rotated label in Fig. 2
\usepackage[usenames,dvipsnames]{color}
 
\title{ Quantization of Binary-Input Discrete Memoryless Channels}

\author{Brian M. Kurkoski and Hideki Yagi, \emph{Members, IEEE} %
\thanks{B.~Kurkoski is with the School of Information Science, Japan Advanced Institute of Science and Technology, Nomi, Japan (email: kurkoski@jaist.ac.jp) }
\thanks{H.~Yagi is with the Dept. of Communication Engineering and Informatics, University of Electro-Communications, Tokyo, Japan (email:  h.yagi@uec.ac.jp).}
\thanks{B.K.~was supported in part by the Ministry of Education, Science, Sports and Culture; Grant-in-Aid for Scientific Research (C) number  23560439.  H.Y.~was supported in part by the Ministry of Education, Science, Sports and Culture; Grant-in-Aid for Young Scientists (B) number 22760270 and JST's Special Coordination Funds for Promoting Science and Technology.}
\thanks{Part of this work was previously presented at the 2010 Information Theory Workshop (ITW) \cite{Kurkoski-itw10}.}
}

\newcommand{\qa}{Quantizer Design Algorithm\xspace}
\newcommand{\qas}{Quantizer Design Algorithm\xspace}

\newcommand{\rX}{\mathsf{X}}
\newcommand{\mX}{\mathcal{X}}
\newcommand{\X}{J}
\newcommand{\x}{x}

\newcommand{\rY}{\mathsf{Y}}
\newcommand{\mY}{\mathcal{Y}}
\newcommand{\Y}{M}
\newcommand{\y}{y}

\newcommand{\rZ}{\mathsf{Z}}
\newcommand{\mZ}{\mathcal{Z}}
\newcommand{\Z}{K}
\newcommand{\z}{z}

\newcommand{\rU}{\mathsf{U}}
\newcommand{\mU}{\mathcal{U}}

\newcommand{\rV}{\mathsf{V}}

\renewcommand{\b}{v}

\renewcommand{\P}{P_{\y | \x}}
\newcommand{\Pq}{P_{\y' | \x}}

\newcommand{\Ppq}{P_{\y' | \x'}}
\newcommand{\Q}{Q_{\z|\y}}
\newcommand{\Qq}{Q_{\z|\y'}}
\newcommand{\T}{T_{\z | \x}}
\newcommand{\Tp}{T_{\z|\x'}}

\newcommand{\p}{p_{\x}}
\newcommand{\pp}{p_{\x'}}

\newcommand{\sx}{\sum_{\x \in \mX}}
\newcommand{\sxp}{\sum_{\x' }}
\newcommand{\sy}{\sum_{\y \in \mY}}
\newcommand{\syq}{\sum_{\y'}}
\newcommand{\sz}{\sum_{\z \in \mZ}}

\newcommand{\A}{{\mathcal A}}
\newcommand{\oT}{{\overline T}}
\newcommand{\oP}{{\overline P}}

\theoremstyle{definition}
\newtheorem{lemma}{Lemma}

\begin{document}

\maketitle

\begin{abstract}
The quantization of the output of a binary-input discrete memoryless channel to a smaller number of levels is considered.   An algorithm which finds an optimal quantizer, in the sense of maximizing mutual information between the channel input and the quantizer output is given.   This result holds for arbitrary channels, in contrast to previous results for restricted channels or a restricted number of quantizer outputs.  In the worst case, the algorithm complexity is cubic $\Y^3$ in the number of channel outputs $\Y$.    Optimality is proved using the theorem of Burshtein, Della Pietra, Kanevsky, and N\'adas for mappings which minimize average impurity for classification and regression trees.  
\end{abstract}


\section{Introduction}

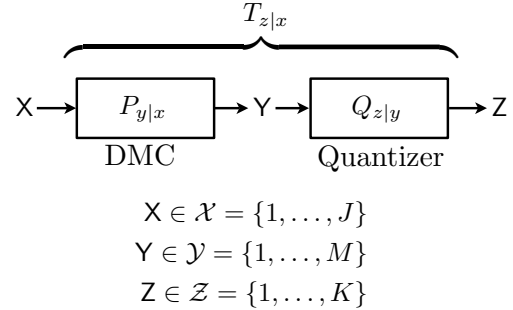
\begin{figure}[t]
%\begin{center}
\hspace{1.1cm} %centering breaks things
\makebox[0pt][l]{\includegraphics[width=6cm]{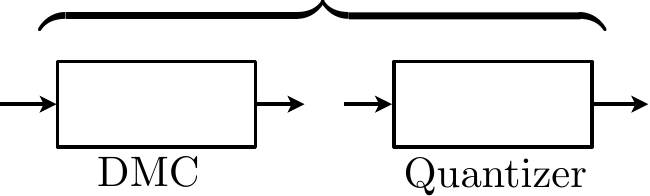}}
\setlength{\unitlength}{1pt}
\begin{picture}(1,1)
\put(-15,21){$\rX$}
\put(24,22){$\P$}
\put(75,21){$\rY$}
\put(112,22){$\Q$}
\put(165,21){$\rZ$}
\put(71,57){$\T$}
\end{picture}
\begin{eqnarray*}
\rX \in \mX = \{1,\ldots,\X \} & & 
\\
\rY \in \mY = \{1,\ldots,\Y \} & &
\\
\rZ \in \mZ = \{1,\ldots,\Z \}
\end{eqnarray*}
%\end{center}
\caption{A discrete memoryless channel followed by a quantizer.   Given $\P$ and $\Z$, find $\Q$ which maximizes $I(\rX;\rZ)$. When the input $\rX$ is binary, this paper gives an algorithm which finds an optimal $Q^*$ with complexity $M^3$.}
\label{fig:fig1}
\end{figure}

Consider a discrete memoryless channel (DMC) with a quantized output, as shown in Fig.~\ref{fig:fig1}. Let the channel input $\rX$ take values from $\mX$, with input distribution $\p$,
\begin{eqnarray*}
\p &=& \Pr(\rX = \x).
\end{eqnarray*}
Let the channel output $\rY$ take values from $\mY$, with channel  transition probabilities $\P$,
\begin{eqnarray*}
\P &=& \Pr(\rY=\y | \rX=\x).
\end{eqnarray*}

The channel output is quantized to $\rZ$, which takes values from $\mZ$, by a possibly stochastic quantizer $\Q$,
\begin{eqnarray}
\Q &=& \Pr(\rZ = \z | \rY = \y),
\end{eqnarray}
so that the conditional probability distribution on the quantizer output $\rZ$ is $\T$,
\begin{eqnarray}
\T &=& \Pr(\rZ = \z | \rX = \x) =  \sy \Q \P.
\end{eqnarray}
Here,  $\mX$, $\mY$ and $\mZ$ are finite sets $\mX =\{1,\ldots,\X \}$,  $\mY =\{1,\ldots,\Y \}$ and  $\mZ=\{1,\ldots,\Z \}$.

The mutual information between $\rX$ and $\rZ$ is:
\begin{eqnarray}
I(\rX;\rZ) &=&  \sz \sx \p \T \log \frac{\T}{\sxp \pp \Tp } \label{eqn:mi}.
\end{eqnarray}
Mutual information $I(\rX;\rZ)$ is convex (lower convex) in $\T$, for fixed $\p$.  Similarly, it is concave (upper convex) in $\p$ for 
fixed  $\T$ \cite[Theorem 2.7.4]{Cover-1991}.  Logarithms are base 2.

The main contribution of this paper is a \qa which finds a mutual information-maximizing quantizer for binary-input DMCs.   The \qa, an instance of dynamic programming, searches over all quantizers that satisfy a condition on quantizer optimality; this condition will be given as Lemma \ref{lemma:nc}. Note that $\Z<\Y$ is of interest, since $\Z \geq \Y$ implies no loss in mutual information due to quantization.   The main result is stated concisely as follows.   

\textbf{Theorem.} The set of all quantizers, including stochastic quantizers, is denoted by $\mathcal Q$.  For an arbitrary binary-input DMC,
the \qas finds a quantizer $Q^*$ which maximizes the mutual information between $\rX$ and $\rZ$:
\begin{eqnarray}
Q^* = \arg \max_{Q \in \mathcal Q} I(\rX;\rZ) \label{eqn:theorem},
\end{eqnarray}
and this algorithm has complexity proportional to $\Y^3$, in the worst case.

This maximization (\ref{eqn:theorem}) is a concave optimization problem. 
General concave optimization is an NP-hard problem \cite{Pardalos-siam86}.  For the specific problem of channel quantization, a naive approach of searching over all $\Z^\Y$ deterministic quantizers has exponential complexity (it will be shown that there exists an optimal quantizer which is deterministic).
The significance of the Theorem is that an optimal quantizer may be found with complexity cubic in $\Y.$   

Discrete channel quantization is shown to be related to the design of classification and regression trees.  Burshtein, Della Pietra, Kanevsky, and N\'adas gave conditions under which  an optimal classification forms a convex set \cite{Burshtein-Annstat92}.  
In this paper, this result is applied to find conditions on optimal channel quantization, which are in turn used to prove the optimality of the \qa.

The flow of the remainder of this paper is as follows. 
Sec.~\ref{sec:relatedwork} gives the motivation, outlines the connections with statistical learning theory, and describes prior work on channel quantization with information-theoretic measures.   Sec.~\ref{sec:conditions} sketches the main argument, giving a condition on an optimal quantizer, while proofs are in the Appendix.   Sec.~\ref{sec:algorithm} gives the \qa which exploits this condition; complexity is also discussed.  A numerical example of quantizing an AWGN channel is given.  Finally, we engage in some discussion in Sec.~\ref{sec:discussion}.

\section{Relationship with Prior Work \label{sec:relatedwork}}

\subsection{Motivation}

The problem of finding good channel quantizers is of importance since most communications receivers convert physical-world analog values to discrete values.  It is these discrete values that are used by subsequent filtering, detection and decoding algorithms.  Since the complexity of  circuits which implement such algorithms increases with the number of quantization levels, it is desirable to use as few levels as possible, for some specified error performance.  

Channel capacity is the maximization of mutual information, so a reasonable metric for designing channel quantizers is to similarly maximize mutual information between the channel input and the quantizer output.   For a memoryless channel with a fixed input distribution, the quantizer which maximizes mutual information will give the highest achievable communications rate.   

In addition to the problem of quantizing a physical channel, this paper's result has other applications.   The original motivation was the implementation of finite-message-alphabet LDPC decoders \cite{Kurkoski-globe08}.   Also, we have recently discovered that discrete channel quantization may have application in the construction of polar codes \cite{Tal-arxiv11}.

\subsection{Learning Theory Connections}

A connection between discrete channel quantization and the area of statistical learning theory can be established.   Classification, which is a type of supervised learning problem, deals with a variable of interest $\rX$ which is stochastically linked to an observation\footnote{Statistical learning literature reverses these, using $\rY$ as the variable of interest, and $\rX$ as the observation, since the input to the classifier is $\rX$, and the output of the classifier is an estimate of $\rY$.} $\rY$ \cite{Bishop-2006}.  This observation must then be classified, or mapped to an estimate of the variable of interest $\rZ$, with some degree of goodness.  The mapping function  $Q:\mY \to \mZ$  is called a classifier in the context of learning theory, and is called a quantizer in this paper.   Tree-based classification methods, called classification trees, regression trees, or decision trees, have numerous applications in machine learning and pattern recognition \cite{Chou-patt91}. A typical example is handwriting recognition, where letters and numbers must be estimated from features of observed handwriting.  

A common tree construction procedure is to successively split nodes, beginning with the root node.  A node should be split in such a way to give ``purer'' descendent nodes, in terms of the classifier's ability to form good estimates \cite{Breiman-1984}.   Impurity is a way of measuring the goodness of a split.   In this paper, a tree is not being constructed, but the $\z \in \mZ$ are analogous to the decisions made by  the tree, and we are interested in the impurity $\phi_\z$.  Various impurity functions  have been suggested \cite[Appendix A]{Chou-patt91}.    The impurity function of interest here is:
\begin{eqnarray}
\phi_{\z} &=& - \sum_{\x = 1}^{\X} \oT_{\x|\z} \log \oT_{\x|\z},
\end{eqnarray}
where $\oT_{\x|\z} = \Pr( \rX = \x | \rZ = \z)$.   This is conditional entropy given $\rZ = \z$, that is  $\phi_{\z} = H(\rX | \rZ = \z)$.   Classification conditioned on $z$ is good if it represents $\rX$ with as few bits as possible; if $\phi_\z$ is zero, then no bits are required, that is, $\rX$ is known exactly.   

The average impurity of a mapping $Q$ is $\Phi(Q)$:
\begin{eqnarray}
\Psi(Q) &=& \sum_{\z = 1}^{\Z} \Pr( \rZ = \z) \phi_\z \\ &=& H(\rX | \rZ).
\end{eqnarray}
Minimization of $\Psi(Q)$ is often of interest in learning theory. 
On the other hand, mutual information, the objective function in \eqref{eqn:theorem}, can be written as:
\begin{eqnarray}
I(\rX ; \rZ) &=& H(\rX) - H(\rX | \rZ). \label{eqn:entropy}
\end{eqnarray}
Thus, minimization of average impurity $\Phi(Q)$ over $Q$ is equivalent to maximization of mutual information over $Q$, since $H(\rX)$ is fixed.  This is the connection between learning theory and discrete channel quantization. 

Burshtein et al.~gave an elegant and general result for a broad class of impurity functions, showing that the preimage which minimizes average impurity forms a convex set in a suitable Euclidean space of observed values \cite{Burshtein-Annstat92}. Since conditional entropy is a member of this class of impurity functions, we can use this result to prove the Theorem.   Note that Coppersmith, Hong and Hosking \cite{Coppersmith-dmkd99}, considering a restricted setting, arrived at a similar conclusion, stating that the preimage of two distinct quantizer outputs are separated by a hyperplane, in a similar Euclidean space.   In the present paper we follow Burshtein et al.

Algorithms for classification can be used for channel quantization.  If optimality is not required, then Chou's iterative algorithm based on the K-means algorithm can find good quantizers $Q$ \cite{Chou-patt91}.  An optimal classifier can be found with complexity $\Y^\Z$ \cite{Burshtein-Annstat92}, which is lower than $\Z^\Y$  for the naive approach, for many cases of interest.

\subsection{Superficially Related Problems}

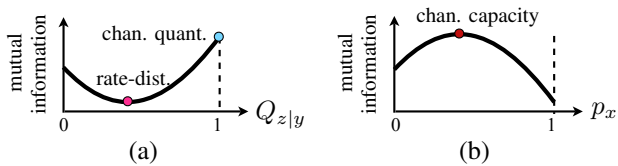
\begin{figure}[t]
\hspace{0.8cm} %centering breaks things
\makebox[0pt][l]{\includegraphics[width=7cm]{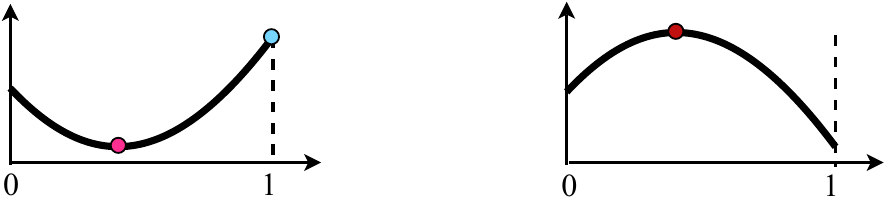}}
\setlength{\unitlength}{1pt}
\begin{picture}(1,1)
{\footnotesize
\put(-26,7){\begin{sideways}\parbox[c]{4cm}{\ \ \ mutual\\ information} \end{sideways}}
\put(99,7){\begin{sideways}\parbox[c]{4cm}{\ \ \ mutual\\ information} \end{sideways}}
\put(5,18){ rate-dist.}
\put(10,35){chan.~quant. }
\put(126,42){ chan.~capacity}
}
\put(68,5){$\Q$}
\put(195,7){$\p$}
\put(20,-10){(a)}
\put(145,-10){(b)}
\end{picture}
\\
\caption{Simplified illustration of information extrema problems.  (a) Mutual information is convex in $\Q$. Rate-distortion computation finds the minimum.   For the present problem, find the maximum.  (b) Mutual information is concave in $\p$; channel capacity computation finds the maximum.}
\label{fig:convex}
\end{figure}

Superficially, the channel quantization problem \eqref{eqn:theorem}  resembles various information-theoretic optimization problems, particularly the computation of the rate-distortion function,
\begin{eqnarray}
R(D) &=& \min_{Q \in \mathcal Q}  I(\rY;\rZ),
\end{eqnarray}
but it is distinct.  Mutual information is convex in $Q$, and for the computation of $R(D)$, 
mutual information is minimized, so this is a \emph{convex} optimization problem \cite[Sec. 13.7]{Cover-1991}, where the distortion constraint is omitted for clarity.   On the other hand, the channel quantization problem is to maximize mutual information in $Q$, leading to a considerably different \emph{concave} optimization problem.    Similarly, computation of the DMC capacity is a convex optimization problem, since mutual information, concave in {$\p$,} is maximized.  The relationship among the rate-distortion problem, the DMC capacity, and the problem treated in this paper is illustrated in Fig.~\ref{fig:convex}.    
Arimoto \cite{Arimoto-it72} and Blahut \cite{Blahut-it72} gave the well-known algorithm to compute the channel capacity and  the rate-distortion function.

Another information extremum problem is the information bottleneck method, from the field of machine learning \cite{Tishby-allerton99}.   The problem setup is identical, using the same Markov chain $\rX \rightarrow \rY \rightarrow \rZ$.  However, this is an information minimization problem, using a Lagrange multiplier to sweep a kind of rate-distortion curve.  Moreover, it is a convex optimization method, using alternating minimization.

\subsection{Information-Theoretic Quantizer Design Criteria}

Information-theoretic measures for channel quantization have been of interest since the 1960s, when Wozencraft and Kennedy suggested using the channel cutoff rate as an optimization criterion \cite{Wozencraft-it66} \cite[Sec. 6.2]{Wozencraft-1990}.  Massey gave a quantizer design algorithm for the binary-input AWGN channel \cite{Massey-isdc74} and Lee extended these results to continuous channels with non-binary inputs \cite{Lee-it76}.   

However, since the channel capacity, which is above the channel cutoff rate, can now be practically approached with turbo codes and LDPC codes, mutual information is a more appropriate measure.   The earliest work we are aware of is the 2002 conference paper of Ma, Zhang, Yu and Kavcic,  which considered quantization of the binary-input AWGN channel \cite{Ma-isita02}.  For the special case of three quantizer outputs, it is straightforward to select a single parameter which maximizes mutual information.    However, for a larger number of outputs, local optimization is feasible and this has higher mutual information than uniform quantization \cite{Liveris-globe03}. 

Singh, Dabeer and Madow considered the problem of  jointly finding capacity-achieving input distributions and AWGN channel quantizers \cite{Singh-com09}.   Again, for an AWGN channel quantized to three levels, optimization over a single parameter was done, but for a larger number of outputs, a local optimization algorithm was used.  In fact, this is a concave-convex problem, and global optimization appears difficult.

Maximization of mutual information has been considered in various application-centric quantization problems.  Channels with memory can be quantized using the information bottleneck method, and quantizers with memory have a higher information rate than memoryless quantizers \cite{Zeitler-com12}.   For flash memories, maximizing mutual information for pulse-amplitude modulation used in flash memories can improve  the performance  of LDPC codes \cite{Wang-globe11}.

\subsection{Contributions of This Paper}

Previous work, discretizing continuous-output channels, either showed optimality in special cases, or considered locally optimal algorithms.   The special cases are restricted to two- and three-level quantization of symmetrical AWGN channels.   The locally optimal quantization algorithms appear to be effective, but they have not been proven optimal.  

By considering discrete-output channels rather than continuous-output channels, further progress can be made on this problem.   In contrast to previous work, the results in this paper hold for arbitrary channels, and for an arbitrary number of quantization levels.  Thus, we believe that this is the first result on globally optimal quantization of general binary-input channels.       Of course, a continuous output channel can be approximated with arbitrarily small discrepancy by a finely quantized channel, which may be used as the original channel $\P$ in this paper.  

While the algorithm assuming uniformly-distributed inputs appeared in conference proceedings previously \cite{Kurkoski-itw10},  contributions of this paper include the proof of its optimality and an extension to non-uniform input distributions, as well as showing the usefulness of statistical learning theory for channel quantization.

\section{ The Structure of Optimal Quantizers  \label{sec:conditions}}

This section develops the condition on an optimal quantizer for binary-input channels.   First, channel quantization is stated as a concave optimization problem, and it is shown that there exists an optimal quantizer that is deterministic.  
Then, a separating hyperplane condition for general number of inputs $J$ is given.
Finally, this condition is specialized to the $J=2$ binary-input discrete memoryless channel case, stated as Lemma~\ref{lemma:nc}.  Proofs are in the Appendix.

\subsection{Restriction to Deterministic Quantizers }

Concave optimization, also known as concave programming or concave minimization and related to global optimization, is a class of mathematical programming problems which has the general form: 
\begin{eqnarray}
\min f(\mathbf x), \textrm{ subject to } \mathbf x \in S, \label{eqn:globaloptimial}
\end{eqnarray}
where $S  \subseteq \mathbb{R}^n$ is a feasible region and $f(\mathbf x)$ is a concave function \cite{Horst-1995} \cite{Horst-1995*2}.    

Mutual information \eqref{eqn:mi} is convex in $\Q$ as well as $\T$, as shown in the Appendix, part \ref{app:prooftwo}.
Thus, the optimization problem given by (\ref{eqn:theorem}), is maximization of a convex function, expressed as a concave programming problem over the $\Y \cdot \Z$ variables $\Q$:
\begin{displaymath}
\max \sx \sz \p (\syq \Qq \Pq)   \log \frac{ \syq \Qq \Pq }{\sxp  \pp  \syq \Qq \Ppq }
\end{displaymath}
\begin{eqnarray}
& & \!\!\!\!\!\!\!\!\!\!  \textrm{subject to:}  \label{eqn:problemtwo}
\\
& & \sz \Q = 1, \ \ \textrm{ $\y=1,\ldots,\Y$ and }\nonumber
\\
& & \Q \geq 0, \ \ \textrm{ $\y=1,\ldots,\Y$ and $\z =1,\ldots,\Z$}. \nonumber
\end{eqnarray}
The constraint enforces that $\Q$ is a conditional probability distribution.  Using a well-known result from concave optimization, the maximum of mutual information is achieved by a deterministic quantizer:

\begin{lemma}
For any DMC and any $\Z$, a deterministic quantizer $Q^*$ maximizes mutual information.  That is, $\Q^* \in \{0,1\}$, for all $\y$ and $\z$. \label{lemma:two}
\end{lemma}

The proof is in the Appendix, part \ref{app:prooftwo}.

As an example to illustrate Lemma \ref{lemma:two}, consider the binary, symmetric errors and erasure channel, with the transition matrix:
\begin{eqnarray}
P &=& \left[ 
\begin{array}{ccc}
1-p-q & p & q \\
q  & p  & 1-p-q
\end{array}
\right],
\end{eqnarray}
for $p,q \geq 0$ and $p+q \leq 1$.
Suppose the three outputs, called 0, erasure and 1, are to be quantized to two levels.  One might expect that symmetry should be maintained by mapping the erasure symbol to the two output symbols with probability 0.5 each.   
However, as Lemma \ref{lemma:two} shows, there exists a deterministic quantizer (which maps the erasure to either 0 or 1) which maximizes mutual information, and this quantizer lacks symmetry between the channel input and quantizer output. For this particular channel, a stochastic quantizer has strictly lower mutual information. 
Note that for the low-SNR AWGN channel quantized to two levels, asymmetric quantizers are optimal as well \cite{Koch-it13}. 

Previous work on the continuous output channel has shown that dithered quantization has no advantage over fixed, deterministic quantizers \cite{Singh-icc07}.  Lemma \ref{lemma:two} extends this idea to DMCs, showing that purely stochastic quantizers, that is, non-deterministic quantizers, never have better performance than deterministic quantizers.

\subsection{Separating Hyperplane Condition for Optimality}

Due to Lemma \ref{lemma:two}, attention may now be restricted to deterministic quantizers $Q$:
\begin{eqnarray}
& Q: \{1,2,\ldots,\Y \} \rightarrow \{1,2,\ldots,\Z \}.
\end{eqnarray}
Let $\mathcal A_z \subseteq \mY$ denote the preimage of $\z$ (in some contexts, the preimage may also be written as $Q^{-1}(z)$). The sets $\A_{\z'}$ and $\A_{\z''}$ are disjoint for $\z' \neq \z''$, and the union of all the sets is $\mY$.

Let $\oP_{\x|\y}$ be the conditional probability distribution on the channel input:
\begin{eqnarray}
\oP_{\x|\y}&=& \Pr(\rX = \x | \rY = \y),
\end{eqnarray}
which depends on the input distribution $\p$ as well as the DMC $\P$.
Each channel output $\y$ corresponds to a vector $\b_\y$:
\begin{eqnarray}
\b_\y &=& \Big[\  \oP_{1 | \y} ,\oP_{2 | \y} ,\ldots, \oP_{\X-1| \y}\  \Big],  \label{eqn:vectorSpace}
\end{eqnarray}
with $\b_\y \in \mU = [0,1]^{\X-1}$,
and for a given channel, the set of all vectors is $\{\b_1, \b_2, \ldots, \b_\Y\}$.   

Define an equivalent quantizer $\widetilde Q$ on the quantization domain $\mU$:
\begin{eqnarray}
\widetilde Q :  \{\b_1, \ldots, \b_\Y\} \rightarrow \{1,2,\ldots,\Z \}.
\end{eqnarray}
The two quantizers are equivalent in the sense that $Q(\y) = \widetilde Q(v_\y) = \z$.  The advantage of the new quantizer $\widetilde Q$ is that the points $v_\y$ exist in the Euclidean space $\mU$, and in that Euclidean space, the following lemma holds.

\begin{lemma}
There exists an optimal quanitzer  $\widetilde Q^*$ for which
any two distinct preimages $\widetilde Q^{*-1}(\z)$ and $\widetilde Q^{*-1}(\z')$ are separated by a hyperplane in $\mU$. %, for $\z, \z' \in \mZ$.
\label{lemma:bur}
\end{lemma}

Burshtein et al.~\cite[Theorem 1]{Burshtein-Annstat92} showed that there must be at least one optimal quantizer for which all preimages are convex in $\mU$.   In the context of discrete channel quantization, convexity corresponds to the condition that the convex hulls of distinct preimages do not intersect, or distinct preimages are separated by a hyperplane.
The application of \cite{Burshtein-Annstat92} is not immediately obvious, and further details and proof are given in the Appendix, part \ref{app:bur} and part \ref{app:burproof}, respectively.

\subsection{Binary Input $\X=2$ Case}

Specializing Lemma \ref{lemma:bur} to the case of $\X=2$, the channel outputs $\b_\y = \oP_{1|\y}$ are in a one-dimensional space.
The separating hyperplane is a point, and an optimal preimage is those points on a line segment.  In order to simplify finding such points,   
it is convenient to assume that the outputs are labeled in such a way that $\oP_{1|\y'} < \oP_{1|\y''}$ implies that $\y' < \y''$, for any $\y' \neq \y''$, that is:
\begin{eqnarray}
\oP_{1|1} < \oP_{1|2} < \cdots < \oP_{1|M}. \label{eqn:bsort}
\end{eqnarray}
There is no loss of generality because the outputs can be re-labeled such that this condition holds.   The benefit should be clear: if  \eqref{eqn:bsort} holds, then for an optimal quantizer $\widetilde Q^*$, the set $\mathcal A_\z$ %$Q^{-1}(\z)$ 
will be a contiguous set of integers. That is, attention may be further restricted to: 
\begin{eqnarray}
\mathcal A_{\z} &=& \{ a_{\z-1}+1, a_{\z-1}+2, \ldots, a_{\z}-1, a_{\z} \} \label{eqn:condition}
\end{eqnarray}
for $\z \in \mathcal Z$, with  $a_0 = 0 $ and $a_{\z-1} < a_{\z}$  and $a_{\Z} =  \Y$.  
The $a_i$ are quantizer boundaries.

The above reasoning proves the following Lemma \ref{lemma:nc}, which uses a slightly more intuitive sorting using log-likelihood ratios to describe a condition satisfied by an optimal quantizer.   Inequalities \eqref{eqn:sortold} are easily obtained from \eqref{eqn:bsort}, and the monotonicity of the $\log$ function is used.

\begin{lemma}
Consider a binary-input DMC that satisfies,
\label{lemma:nc}
\begin{eqnarray}
\log \frac{P_{1|1}}{P_{1|2}} < \log \frac{P_{2|1}}{P_{2|2}} < \cdots < \log \frac{P_{M|1}}{P_{M|2}}. \label{eqn:sortold}
\end{eqnarray}
Then, there exists an optimal quantizer $Q^*$, such that each $\A_z^*$ consists of a contiguous set of integers, and so the optimal quantizer boundaries $a^*_{\z}$  satisfy:
\begin{eqnarray}
& 0 < a^*_1 < a^*_2 < \cdots < a^*_{\Z-1} < \Y. \label{eqn:sort}
\end{eqnarray}
\end{lemma}
Note that the condition \eqref{eqn:sortold} does not depend on the input distribution $\p$.

The necessity of the contiguousness condition \eqref{eqn:sort} in Lemma \ref{lemma:nc} for an optimal quantizer to be deterministic cannot be shown using Lemma \ref{lemma:two} 
alone.  However, the more restricted classification setting of Coppersmith et.~al \cite{Coppersmith-dmkd99} can be applied here to demonstrate the necessity the condition.  The relevance of necessity to the \qas is discussed in the next section.

Strict inequalities are used in \eqref{eqn:bsort} and \eqref{eqn:sortold} because if  
\begin{eqnarray}
\frac{P_{\y|1}}{P_{\y|2}} &=& \frac{P_{\y+1|1}}{P_{\y+1|2}}, \label{eqn:equiv_outputs}
\end{eqnarray}
then outputs $\y$ and $\y+1$ can be combined to a single output with the likelihood $P_{\y|\x}+P_{\y+1|\x}$ for input $\x$ to form a new channel with $\Y-1$ outputs.  The likelihood ratio for the combined output, 
\begin{eqnarray}
\frac{P_{\y|1}+P_{\y+1|1}}{P_{\y|2}+P_{\y+1|2}} \label{eqn:new_output}
\end{eqnarray}
is equal to (\ref{eqn:equiv_outputs}). 
The new channel is operationally equivalent to the original channel, and moreover the two channels have the same mutual information.

\section{\qa \label{sec:algorithm} }

This section describes the \qa which finds a quantizer $Q^*$ that maximizes mutual information $I(\rX;\rZ)$ over quantizer boundaries satisfying  $a_1 < a_2 < \ldots < a_{\Z-1}$. By Lemma \ref{lemma:nc}, $Q^*$ provides the maximum mutual information over all quantizers.   The \qa, as well as Lemma \ref{lemma:nc}, are restricted to binary-input DMCs.

\subsection{Partial Mutual Information}

\newcommand{\syA}{\sum_{y \in \mathcal{A}_{\z}}}
\newcommand{\syAA}{\sum_{y' \in \mathcal{A}_{\z}}}
\newcommand{\sya}{\sum_{y = a'+1}^{a}}
\newcommand{\syaa}{\sum_{y' = a'+1}^{a}}

A partial sum of mutual information is called ``partial mutual information."   Partial mutual information $\iota$ is the contribution that one or more quantizer outputs makes to the total mutual information.  Consider that the conditional probability distribution on the quantizer output $\T$  can be written as:
\begin{eqnarray}
\T &=& \syA \P = \sum_{y = a_{\z-1}+1}^{a_\z} \P.
\end{eqnarray}

The contribution that quantizer output $\z$ makes to mutual information is called the partial mutual information $\iota_{\z}$:
\begin{eqnarray}
\iota_{\z}  &=&  \sx \p (\syA \P ) \log \frac{\syA \P }{\sxp \pp \syAA P_{\y' | \x'}}, \nonumber
\label{eqn:pmim}
\end{eqnarray}
so total mutual information is the sum of all the partial mutual information terms:
\begin{eqnarray}
I(\rX;\rZ) &=& \sz  \iota_{\z}.
\end{eqnarray}

For one quantizer output with boundaries $a', a$ with $a' \leq a$, denote the partial mutual information by  $\iota(a' \rightarrow a)$, which is:
\begin{align}
\iota(a' \rightarrow a) &=& \!\!\!\! \sx \p \sya \P \log \frac{\syaa \Pq }{\sxp \pp \syaa P_{\y' | \x'}},
\nonumber \\ & & \label{eqn:g}
\end{align}
So if $\A_\z = \{a_{\z-1}+1,\ldots, a_\z\}$, then $\iota_\z = \iota(a_{\z-1} \rightarrow a_\z)$.

\begin{figure}[t]
\begin{center}
\includegraphics[width=7cm]{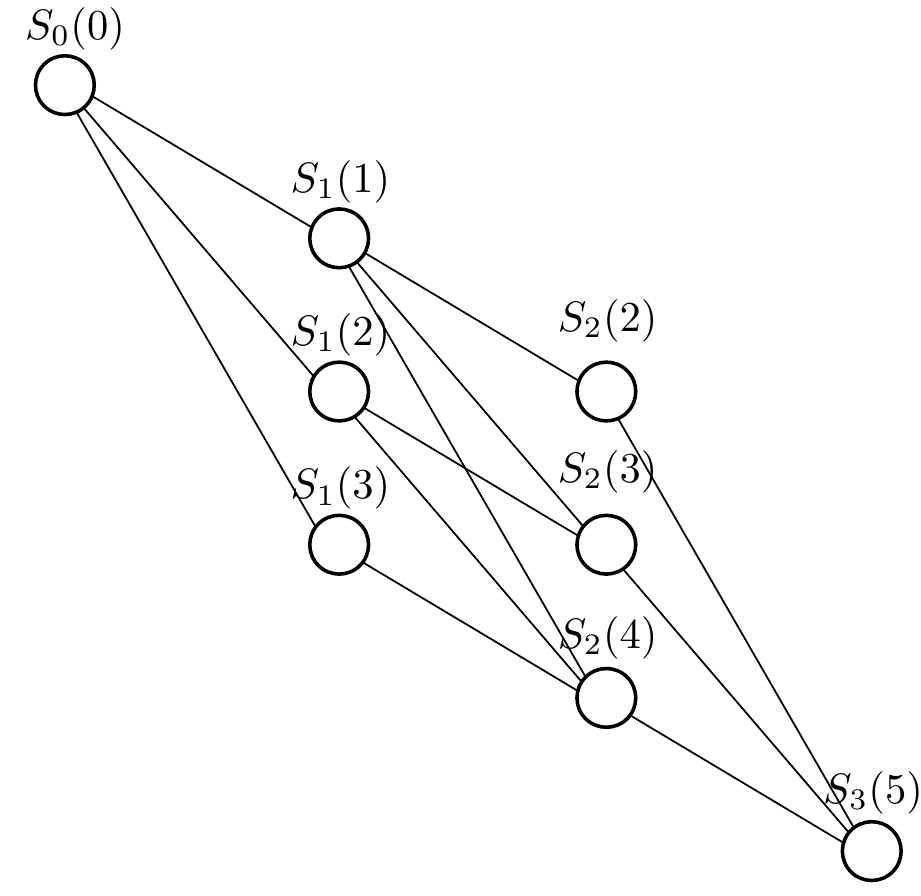}
\end{center}
\caption{Trellis-type illustration showing the relationship between state values $S_\z(a)$, for $\Y=5$ and $\Z=3$.}
\label{fig:StateDiagram}
\end{figure}

\subsection{\qa}

The \qa is an instance of dynamic programming.   The algorithm has a state value $S_\z(\y)$, which is the maximum partial mutual information when channel outputs 1 to $\y$ are quantized to quantizer outputs 1 to $\z$.  This can be computed recursively by conditioning on the state value at time index $\z-1$:
\begin{eqnarray}
S_\z(a) &=& \max_{a'} \Big( S_{\z-1}(a') + \iota(a' \rightarrow a) \Big), \label{eqn:recursion}
\end{eqnarray}
where the maximization is taken over $a' \in \{\z-1,\ldots, a-1\}$.  Clearly, $S_\Z(\Y)$ is the maximum total mutual information.   The sequence $S_0(0), S_1(a_1), \ldots, S_\Z(a_\Z)$ which gives the maximum of total mutual information corresponds to the optimum quantizer whose boundaries are $\{a_1, a_2, \ldots, a_\Z\}$.  The relationship between the state values is illustrated in a trellis-type diagram in Fig.~\ref{fig:StateDiagram}, for $\Y=5$ and $\Z=3$.  

\emph{\qas}   
\begin{enumerate}
  \item Inputs: 
  \begin{itemize}
    \item Binary-input discrete memoryless channel $\P$.  If necessary, modify labels to satisfy (\ref{eqn:sortold}).
    \item The number of quantizer outputs $\Z$.
  \end{itemize}
  \item Initialize $S_0(0) = 0$.  
  \item Precompute partial mutual information.  For each $a' \in \{0,1, \ldots, \Y-1\}$ and for each $a \in \{a'+1, \ldots, t \}$ (where $t = \min\big( a'+1+\Y-\Z, \Y\big)$): 
  \begin{itemize} \item  compute $\iota(a' \rightarrow a)$ according to (\ref{eqn:g}). \end{itemize}
  \item Recursion. For each $\z \in \{ 1,\ldots,\Z \}$, and for each $a \in \{ \z,\dots,\z+\Y-\Z \} $, 
  \begin{itemize}
    \item compute $S_\z(a)$ according to (\ref{eqn:recursion}),
    \item store the local decision ${h}_\z(a)$:
    \begin{eqnarray*}
   {h}_\z(a) =\arg \max_{a'} S_{\z-1}(a') + \iota(a' \rightarrow a),
    \end{eqnarray*} 
    where the maximization is taken over $a' \in \{\z-1,\ldots, a-1\}$.
  \end{itemize}
  \item Find an optimal quantizer by traceback.   Let $a^*_\Z = \Y$.   For each $\z \in \{ \Z-1,\Z-2,\ldots,1\}$:
  \begin{eqnarray*}
  a^*_\z = {h}_{\z+1}(a^*_{\z+1}) .
  \end{eqnarray*}
  \item Outputs:
  \begin{itemize}
    \item The optimal $a^*_1, a^*_2, \ldots, a^*_{\Z-1}$.   Equivalently, output the matrix $Q^*$, where row $\z$ of $Q^*$ has ones in columns $a_{\z-1}+1$ to $a_\z$ and zeros in all other columns.   
    \item The maximum mutual information, $S_\Z(\Y)$.
  \end{itemize}
\end{enumerate}

There may be multiple optimal quantizers.  To find these, a tie-preserving implementation should collect all locally optimal decisions and tracebacks to produce multiple optimal quantizers.  This was not explicitly indicated, to keep the notation simple.

With arbitrary tie-breaking, Lemma \ref{lemma:nc} guarantees that the \qas will find one of the optimal quantizers.  However,  \eqref{eqn:sort}  in Lemma \ref{lemma:nc} can be shown to be a necessary condition for optimality, using \cite{Coppersmith-dmkd99}.  Thus, a tie-preserving implementation will find all optimal deterministic quantizers.  This can be further extended to show that probabilistic quantizers are suboptimal; this is done by showing the strict inequalities $\oP_{1|1} < \oP_{1|2} < \cdots < \oP_{1|M}$ in  \eqref{eqn:bsort} lead to strict convexity of mutual information.  In summary, this paper has shown in detail that the \qas will find at least one optimal quantizer, and has given a sketch of an argument that the tie-preserving implementation finds all optimal quantizers.

\begin{figure}[t]
\begin{center}
\includegraphics[width=8cm]{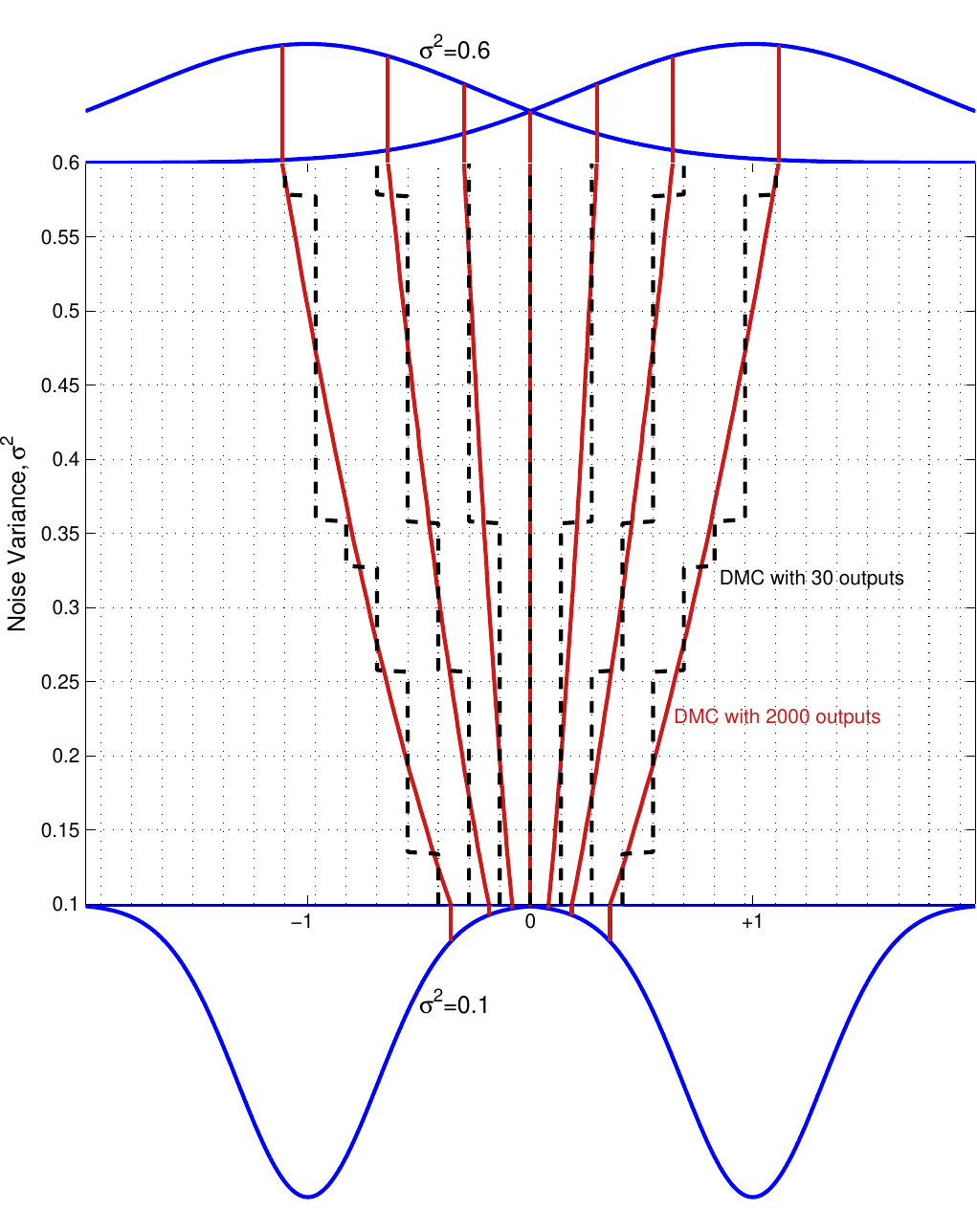}
\end{center}
\caption{Optimal quantization to $\Z=8$ levels of a DMC derived from a finely-quantized binary-input AWGN channel, with noise variance $\sigma^2$.  Solid and dotted lines show quantization boundaries when the AWGN channel is quantized to $\Y=2000$ and $\Y=30$ levels, respectively.}
\label{fig:awgnquant}
\end{figure}

\begin{figure}[t]
\begin{center}
\includegraphics[width=8cm]{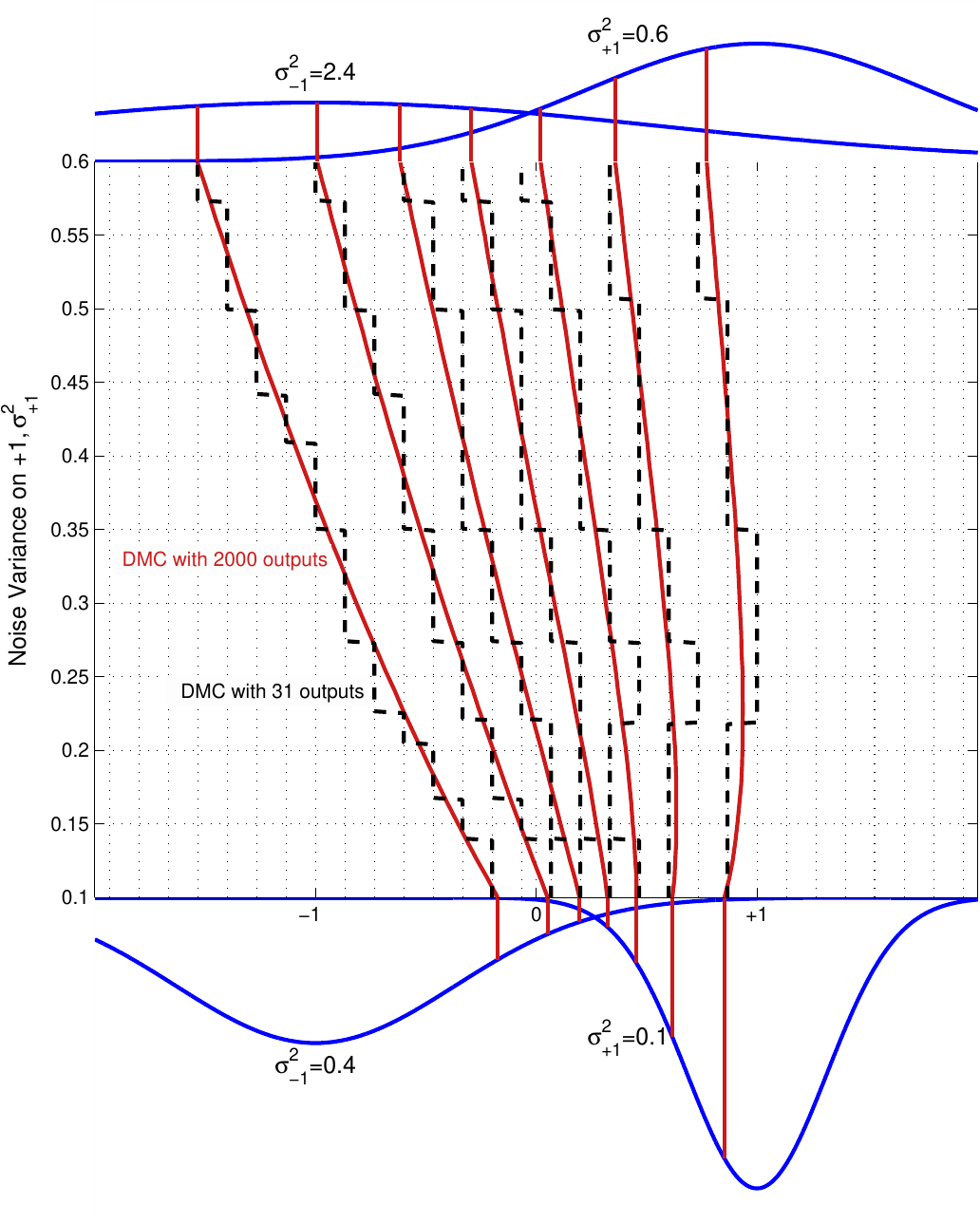}
\end{center}
\caption{Optimal quantization of an asymmetric channel.  Similar to Fig.~\ref{fig:awgnquant}, but the data-dependent noise variance is $\sigma^2_{-1} = 4 \sigma^2_{+1}$. Solid and dotted lines show quantization boundaries when the AWGN channel is quantized to $\Y=2000$ and $\Y=31$ levels, respectively. }
\label{fig:awgnquantasym}
\end{figure}

\subsection{Complexity}

A naive optimization approach for (\ref{eqn:problemtwo}) is to search over all $\Z^\Y$ candidate solutions, which is searching over all deterministic quantizers.  This has complexity exponential in $\Y$.   

Burshtein et al.~suggested searching over the quantizers satisfying Lemma \ref{lemma:bur} \cite{Burshtein-Annstat92}.   
There are ${\Y-1 \choose \Z-1}$ such quantizers, formed by selecting the $\Z-1$ values $a_1$ to $a_{\Z-1}$ from the set of $\Y-1$ integers.    In this case, the complexity grows as $\Y^\Z$.     This is polynomial in $M$ when $K$ is fixed.  However, for the important case of $\frac K M$ fixed, the complexity remains exponential.  

On the other hand, the \qas has polynomial complexity $\Y^3$ in the worst case, and more generally  has complexity $\Z (\Y-\Z)^2$.    The main computational burden is to pre-compute $\iota(a' \rightarrow a)$ in step 3.   Since $a'$ is from a set of size $\Y$ and $a$ is from a set of size at most $\Y-\Z+1$, the number of $\iota$ computations is proportional to $\Y^2$.   Note that in (\ref{eqn:g}), the sum on $\y'$ could be over as many as $\Y-\Z+1$ terms.   However, since this sum can be computed recursively, the complexity remains proportional to $\Y^2$.
Also, for each $\z$ in step 4, roughly $\frac 1 2  (\Y-\Z)^2$ add/compare operations are needed, and there are $\Z$ such steps.       This results in a number of operations roughly  $\frac 1 2 \Z (\Y-\Z)^2$.  For fixed $\Y$, the expression $\Z (\Y - \Z)^2$ has its maximum value at $\Z = \frac{1}{3} \Y$.   The maximum value is $\frac 4 {27} \Y^3$, so the complexity is proportional to $\Y^3$.   

Dynamic programming principles are used to show the optimality of the algorithm, and details are in the Appendix, part \ref{app:theoremproof}.   The complexity result, along with the proof of optimality in the Appendix, proves the Theorem.

\subsection{Example: Finely Quantized Continuous-Output Channel  \label{sec:applications}}

While the \qa can only be applied to discrete channels, it can be used to obtain good coarse quantization of a continuous-output channel, by first uniformly and finely quantizing to $\Y$ levels. This can be illustrated for the binary-input AWGN channel with $\pm 1$ inputs and Gaussian noise variance $\sigma^2$. 

Fig.~\ref{fig:awgnquant} was created by first uniformly quantizing the AWGN channel between $-2$ and $+2$ with $\Y=2000$ or $\Y=30$ steps; the two input symbols are equally likely.  Then, the \qa was applied with $\Z=8$ quantizer outputs.   The figure shows the $\Z-1= 7$ quantization boundaries of the underlying finely quantized DMC.   

Fig.~\ref{fig:awgnquantasym} shows similar quantization for an asymmetric channel, where the Gaussian variance is input dependent, with $\sigma^2_{-1} = 4 \sigma^2_{+1}$.    In this case, the two finely quantized DMCs have $\Y=2000$ and $\Y=31$. Since the channel is asymmetric, the the optimum quantizers are also asymmetric.  It is of interest to see that the center boundary $a_{\frac{\Z}{2}}$ crosses the value 0 as $\sigma^2_{+1}$ increases.  Note also the non-monotonic behavior of the right-most boundary.

\section{Discussion \label{sec:discussion}}

For a binary-input discrete memoryless channel, the main contribution of this paper is an algorithm that finds an optimal quantizer  with cubic complexity $\Y^3$ in the number of channel outputs $\Y$.  Previous results on optimal channel quantization were restricted to symmetrical channels or a small number of quantizer outputs.  The \qas may have various applications beyond the quantization of physical channels; already the design of polar codes and implementation of LDPC decoders have been identified as targets.

A result from statistical learning theory was applied to prove optimality.   Since conditional entropy $H(\rX | \rZ)$ satisfies the conditions of an impurity functions, Burshtein et.~al's theorem on the convexity of the optimal mapping was used \cite{Burshtein-Annstat92}.  

In order to obtain an efficient algorithm, attention was restricted to binary inputs, so that convexity in one dimension corresponded to grouping contiguous integers.  However, the theorem of Burshtein et al., and correspondingly Lemma~\ref{lemma:bur}, is not restricted to binary inputs channels, and could be used to develop quantizer design algorithms for non-binary channels.    Such an extension to a $\X$-input channel requires efficient enumeration sets formed from separating hyperplanes in $\X-1$ dimensional space.   As previously noted \cite{Burshtein-Annstat92}, generalized counting functions \cite{Cover-computers65} give a bound on complexity.   Suboptimal algorithms for tree classification can find good quantizers, if optimality is not required \cite{Chou-patt91}.  However, finding an optimal and efficient quantization algorithm remains a sophisticated problem.   

A natural question is how to find the jointly optimal input distribution and channel quantizer for a given DMC.  We have already considered a simple extension of the quantization algorithm which either finds the jointly optimal input distribution and channel quantizer or declares a failure \cite{Kurkoski-isit12}.   However, this is a convex-concave optimization problem, a class of problems which is difficult, and finding the jointly optimal solution remains an open problem.   The generalization from mutual information to other information-theoretic metrics is also of interest.   While we have already considered the random error exponent \cite{Yagi-icc12} as a generalization of the cutoff rate, the handling of the new metrics is also an open problem.

\section{Acknowledgments} 

The authors extend gratitude to David Burshtein for pointing out the connection between channel quantization and learning theory, and showing the use of  \cite{Burshtein-Annstat92} to obtain straightforward proofs.  

\begin{appendix}

\subsection{Proof of Lemma \ref{lemma:two} \label{app:prooftwo} }

Background on convex optimization and the proof of Lemma \ref{lemma:two} are given.   

First, it is shown that mutual information \eqref{eqn:mi} is convex in $\Q$ as well as $\T$, 
for fixed $\p$ and fixed $\P$.
The relationship $\T  =  \sum_{y} \Q \P$ is an affine transform. 
If a function is convex, then it is also convex in an affine transform of the original arguments \cite[Sec.~3.2]{Boyd-2004}, so mutual information is convex in $\Q$.

Referring to ~\eqref{eqn:globaloptimial}, the following is a well-known result from concave optimization (or, concave minimization):
\begin{lemma} 
\cite[Theorem 1.19]{Horst-1995*2}   A concave (convex) function $f: S \rightarrow \mathbb R$ attains its global minimum (maximum) over $S$ at an extreme point of $S$. \label{lemma:one}
\end{lemma}

If $S$ is a polytope, as in this paper, then the extreme points are its vertices.  There may be multiple local maxima when the objective function is convex.  Lemma \ref{lemma:one} can be visualized in one dimension in Fig.~\ref{fig:convex}-(a), where it is clear that the maximum must be at either endpoint 0 or 1, that is, the vertices of the line segment that is the feasible region.   

To prove Lemma \ref{lemma:two}, observe that the constraints on the variables $\Q$ are given by~\eqref{eqn:problemtwo}.
For any fixed $\y$, the $\Z$ variables $Q_{1|\y}, \ldots, Q_{\Z|\y}$ are restricted to be in the polytope which is a $(\Z-1)$-dimensional simplex:
\begin{eqnarray}
& \Big\{ \Q \in \mathbb R^{\Z} | \sum_{\z=1}^{\Z} \Q = 1, \Q \geq 0\Big\}.
\end{eqnarray}

For the vertices of this polytope,  $\Q=0$ or 1, with $\z = 1,2,\ldots,\Z$.    By Lemma \ref{lemma:one}, there is at least one vertex which obtains an optimal solution, that is, there exists an optimal quantizer which is deterministic. \hfill \qed

Note that if mutual information is strictly convex in $\Q$, then at least one extreme point will have mutual information strictly greater than any stochastic quantizer, that is, all optimal quantizers will be deterministic.

\subsection{Minimum Average Impurity \label{app:bur}}

Burshtein et al.~showed that a classifier which minimizes the average impurity $\Psi$ induces a convexity condition  \cite{Burshtein-Annstat92}.  The main result is given here verbatim but with renamed variables, in order to elucidate the notation.   The proof of Lemma \ref{lemma:bur} is given in the next section.

Let $(\rY,\rU)$ be jointly-distributed random variables with values in $\mY \times \mathbb R^n$, and let $\mU \subset \mathbb R^n$ be the convex hull of the range of $\rU$.  Let $\mathcal Z$ be a finite set $\mathcal Z = \{1,2,\ldots,\Z\}$, and let $\phi : \mZ \times \mU \rightarrow \mathbb R^1$ be concave in its second argument.   

Let $\rV = E[ \rU | \rY ]$ so that $\rV : \mY \rightarrow \mU$ is a random variable on $\mY$, where $E[ \cdot ]$ denotes expectation.   For any measurable partition $Q: \mY \rightarrow \mZ$, define the objective function $\Psi(Q)$ (to be minimized) by:
\begin{eqnarray}
\Psi(Q) &=& E \left[ \phi \Big(Q(\rY),E[\rU|Q(\rY)] \Big)\right]  \\
&=& E \left[ \phi\Big(Q(\rY), E[\rV|Q(\rY)] \Big)\right] .
\end{eqnarray}
Explicitly,
\begin{eqnarray}
\Psi(Q) &=& \sz \Pr(Q^{-1}(\z)) \phi \Big(\z, E[\rU|Q(\rY)=\z] \Big). \label{eqn:generalObjective1}
\end{eqnarray}

Under these conditions, the following holds:
\begin{lemma} 
\cite[Theorem 1]{Burshtein-Annstat92}   For any $Q$, there exists $\widetilde Q: \mU \rightarrow \mZ$ such that $\Psi(\widetilde Q (\rV) ) \leq \Psi(Q)$ and such that $\widetilde Q^{-1}(\z)$ is convex for all $\z \in \mZ$. 
\label{lemma:b}
\end{lemma}

\subsection{Proof of Lemma \ref{lemma:bur} \label{app:burproof}}

To prove Lemma \ref{lemma:bur}, we show that $H(\rX | \rZ)$ is the form of \eqref{eqn:generalObjective1}, and thus minimizing $\Psi(Q)$ is equivalent to maximizing $I(\rX;\rZ)$ as in \eqref{eqn:theorem}. 
Here, 
\begin{eqnarray}
H(\rX | \rZ) &=&  \sz \Pr(\rZ = z) H(\rX | \rZ = z)  \label{eq:conditional_entropy}
\\
 &=& - \sz  \Pr(\rZ = z) \sx \oT_{\x | \z} \log  \oT_{\x | \z} 
\end{eqnarray}
where $\oT_{\x | \z} = \Pr( \rX = \x | \rZ = \z)$.

Define $\rU$ as 
\begin{eqnarray}
\rU &=& [ \delta(\rX,1), \ldots, \delta(\rX,\X-1) ], \label{eqn:rvU}
\end{eqnarray}
 where
\begin{eqnarray}
\delta(\rX,i)  &=& \left\{ \begin{array}{cc}
1 & \textrm{if $\rX=i$} \\
0 & \textrm{otherwise}
\end{array} \right . .
\end{eqnarray}

Then, by denoting $Q(\rY) = \rZ$:
\begin{eqnarray}
E[\rU | Q(\rY) = \z] &=& [ \oT_{1|\z}, \oT_{2|\z}, \ldots, \oT_{\X-1|\z} ].
\end{eqnarray}
Explicitly giving $\phi$ as:
\begin{eqnarray}
\phi( t_1, t_2, \ldots, t_{\X-1} ) &=& - \sum_{i=1}^{\X} t_i  \log t_i,
\end{eqnarray}
where $t_\X = 1 -  (t_1 + \cdots + t_{\X-1}) $. 
We can see:
\begin{eqnarray}
\phi( E[\rU | Q(\rY) = \z ] ) &=& H(\rX | \rZ = \z),
\end{eqnarray}
where dependence of $\phi$ on $z$ is not needed and dropped.  Note that $\Pr(Q^{-1}(\z)) $ is identical to $\Pr(\rZ = z)$, and so $H(\rX | \rZ)$ is expressed in the form of \eqref{eqn:generalObjective1} via \eqref{eq:conditional_entropy}.

Consider now $E[\rU | \rY = \y]$ which explicitly is:
\begin{eqnarray}
E[\rU | \rY = \y] &=& \left[ \oP_{1|\y} , \oP_{2|\y} ,\ldots, \oP_{J-1|\y} \right]. \label{eqn:probabilityvector}
\end{eqnarray}
These are the values $\b_y$ given earlier in \eqref{eqn:vectorSpace}.   Then, $\rV = E[\rU|\rY]$ is a random variable on $\mY$, which takes the value  \eqref{eqn:probabilityvector} with probability $\Pr(\rY = \y)$.  Here, $\rV$ and $\rY$ are random variables, both on the sample space $\mY$.   A quantizer $Q$ defined for $\rY$ is equivalent to a quantizer $\widetilde Q$ for $\rV$.  The space mapped to $\mZ$ is transformed from $\mY$ to $\mU$.   The points $\b_y$ are interpreted as points in a $(J-1)$-dimensional Euclidean space, inside the convex hull $\mU$.   It is on this space that  $\widetilde Q^{-1}(z)$ is convex for all $\z \in \mZ$.

The convexity conditions of Lemma \ref{lemma:b} apply to $\widetilde Q$, the quantizer of $\rV$.   Thus by Lemma \ref{lemma:b}, the set $\widetilde Q^{*-1}(z)$ is convex, for each $\z \in \mZ$, proving Lemma \ref{lemma:bur}. \hfill \qed

\subsection{Proof of Theorem's Optimality Part \label{app:theoremproof} }

In this section, the optimality part of the Theorem is proved.   Recall that by Lemma \ref{lemma:nc} an optimal quantizer satisfies $a_1 < a_2 < \cdots < a_{\Z-1}$, where $a_i$ are the quantizer boundaries.  In the language of dynamic programming, a problem exhibits optimal substructure if the optimal solution contains optimal solutions to subproblems.   If this condition holds, then dynamic programming provides the optimal solution, and moreover, the optimal substructure should be exploited in the optimization \cite[Sec.~15.3]{Cormen-2001}.   

For the Theorem, the subproblem consists of finding the quantizer which maximizes partial mutual information for some partial quantization of the outputs.    In detail, recall $S_k(a)$ is the maximum of partial mutual information when channel outputs 1 to $a$ are quantized to quantizer outputs 1 to $k$,
\begin{eqnarray}
S_k(a) &=& \max_{a'} \Big( S_{k-1}(a') + \iota(a' \rightarrow a) \Big),
\end{eqnarray}
where the maximization is over $a' \in \{k-1,\ldots,a-1\}$.

For fixed $\z$ and $a$, assume that $S_\z(a)$ is the maximum of partial mutual information, corresponding to an optimal quantization of channel outputs 1 to $a$ to the quantizer outputs 1 to $\z$.      Let $\widetilde a$ be the boundary of quantizer output $\z-1$, that is:
\begin{eqnarray}
S_\z(a) &=&  S_{\z-1}(\widetilde a) + \iota(\widetilde a \rightarrow a),
\end{eqnarray}
so that the preimage of quantizer output $\z$ is $\A_\z = \{\widetilde a + 1, \ldots, a\}$.
Then, the quantizer for channel outputs $1$ to $\widetilde a$ must also be optimal.  This is true because  if another quantizer of 1 to $\widetilde a$ produced higher mutual information, then the quantization of 1 to $a$ would also have higher partial mutual information, leading to a contradiction of the assumption that 1 to $a$ was optimally quantized.

Along with the earlier statement that the complexity is proportional to $\Y^3$, the proof of the Theorem is completed. \hfill \qed

\end{appendix}

\bibliographystyle{ieeetr}
\bibliography{/bibtex/abbrev,/bibtex/brian,/bibtex/starbib,/bibtex/briankurkoski-conf,/bibtex/briankurkoski-noreview,/bibtex/briankurkoski-recent}

\newpage

\end{document}